\documentclass[twocolumn,showpacs,preprintnumbers,amsmath,amssymb,floatfix]{revtex4}

\usepackage{graphicx}
\usepackage{dcolumn}
\usepackage{bm}
\begin{document}

\title{\bf Generation and Measurement of Non Equilibrium Spin Currents  in
Two Terminal Systems}
\author{T.~P.~Pareek$^{1}$ and A. M. Jayannavar$^{2}$}
\affiliation{$^{1}$Harish Chandra Research Institute, Chhatnag Road, Jhusi,
Allahabad - 211019, India \\
$^{2}$Institute of Physics, Sachivalya Marg, Bhubaneswar - 751005, India
}%

\begin{abstract}
Generation and measurement of non-equilibrium spin current in two probe configuration is
discussed. It argued and shown that spin current can be generated in two terminal non-magnetic system. Further it is shown that these spin currents can be measured via conductance
in two probe configuration when the detector probe is ferromagnetic. 
\end{abstract}

\maketitle

The generation, manipulation and detection of spin currents, a flow of angular momentum,
is a major goal of spin-based electronics. Thus establishing methods for efficient generation
and detection of spin currents is a key for further advancement of spintronics.
Amongst the different approaches for spin current generation and manipulation, spin-orbit(SO) coupling
is attracting considerable interest. It has been predicted that in a SO coupled system
a non zero spin current flows in a direction perpendicular to the applied electric field 
\cite{dyakonov}. This gives rise to related spin-Hall effect (SHE) causing spin accumulation at
the edges of sample \cite{hirsch,tribhu_prl}. The spin accumulation caused by SHE has been 
detected by optical techniques 
in a semiconductor channel \cite{kato} and in two-dimensional electron gas (2DEG) in
semiconductor heterostructures \cite{wunderlich, awschalom}. Recently electronic detection of spin accumulation 
in Hall cross geometry which is a four probe system, was reported in Ref.\cite{tinkham,kimura}.
Till date most of the theoretical and 
experimental studies are concerned with the observation of SHE which is a spin accumulation caused by the flow
of spin current in multi-probe systems. To the best of our knowledge the possibility of
spin current generation and detection in two probe systems via conductance measurement has not been reported yet.
In this article we address this possibility. We consider two configuration shown in
Fig.~1(a) and 1(b) where a 2DEG with SO coupling is sandwiched between two
ideal leads without SO coupling. In Fig.~1(a) both leads are non-magnetic while
in Fig.~1(b), the detector lead is ferromagnetic(FM) and injector lead is non-magnetic.

\begin{figure}
\includegraphics[width=\linewidth,height=1.9in,angle=0,keepaspectratio]{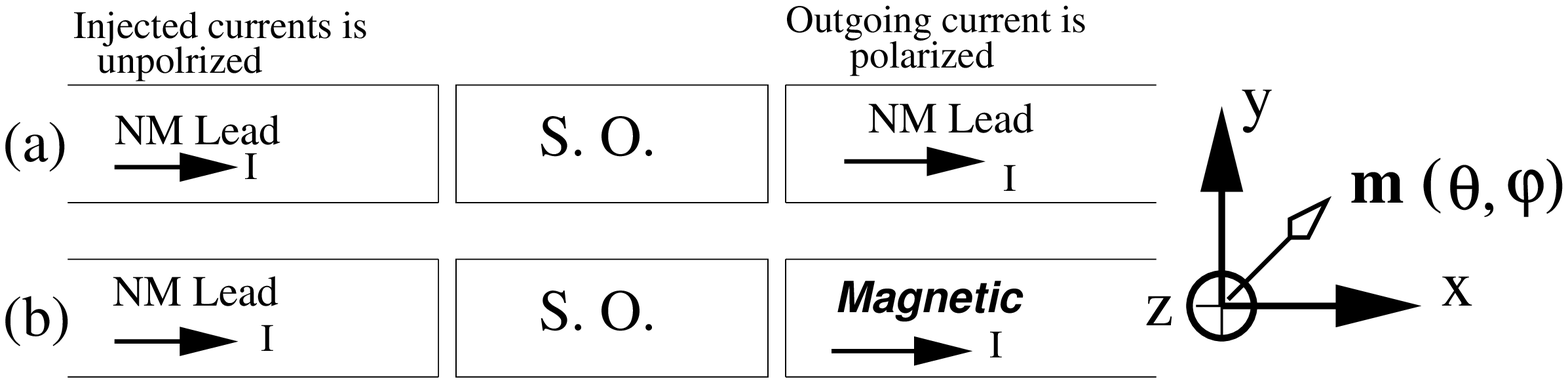}
\caption{\label{fig1} Setup for spin current generation and detection (a) current leads are
non-magnetic and injected currents is unpolarized but outgoing current is polarized
due to SO region (b) Detector lead is magnetic; and the conductance depends on absolute director of magnetization ($\theta$,$\phi$) because of the spin current generated by SO region}
\end{figure}

An intuitive understanding of the non-equilibrium spin current generation can be 
gained by considering the specific case of Rashba SO coupling which in 2DEG has the form,
$H_{R}=\lambda (k_{x} \sigma_{y} - k_{y}\sigma_{x})$, where $k_{x}$ and $k_{y}$ are components of momentum vector
in the plane of 2DEG and  $\sigma_{x}$ and $\sigma_{y}$ are Pauli matrices.
Consider the situation when the 
incident current flows along positive {\it x} direction (Fig. 1(a)).
In this particular case the Rashba interaction is effectively, $\lambda (k_{x} \sigma_{y})$ because
$k_y$=0.
Therfore the outgoing current gets polarized along the {\it y} axis.
In general SO interaction polarizes scattered beam. This 
general result is
valid for arbitrary form of SO interaction viz. Rashba\cite{rashba}, Dresselhaus\cite{dresselhaus} or impurity induced SO interaction \cite{tribhu_prl,tribhu_pramana,tribhu}.
However the direction and magnitude of polarization
for the scattered beam depends on the exact form of SO interaction\cite{landau}.
{\it Therefore the outgoing charge current in non-magnetic lead two is polarized even though the
incident current is unpolarized}.
{\it Hence it is a true flow of non-equilibrium angular momentum,
from SO coupled region to the detector lead}.


Let us now focus on the conceptual problem of non-equilibrium spin current measurement via
conductance in two probe system with ferromagnetic(FM) detector shown in Fig. 1(b). 
According to Onsager's relation the conductance of a two terminal system with one contact 
being ferromagnetic
is invariant upon magnetization reversal, {\it i.e.},G($\hat{\bm{m}}$)=G($-\hat{\bm{m}}$), where $\hat{\bm{m}}$
is unit vector along the magnetization direction. In absence of SO interaction, system shown in
Fig. 1(b) is rotationally invariant, therefore conductance does not depend on the
direction of magnetization and Onsager symmetry is satisfied trivially. In presence of
SO interaction since rotation symmetry is broken therefore conductance 
depends on the
absolute direction of magnetization for configuration shown in Fig.~1(b).
This is because
the detector lead is ferromagnetic (Fig. 1(b)), therefore
electrons polarized in the direction of magnetization will transmit easily compared to the
electrons polarized along another direction. Hence the conductance of NM/SO/FM system (Fig. 1(b))  
will depend 
on the absolute direction of magnetization $\theta$ and $\phi$ and will be symmetric in $\theta$ 
and $\phi$
consistent with Onsager symmetry.In other words,
G($\hat{\bm{m}}_{1}$)$\neq$ G($\hat{\bm{m}}_{2}$), where $\hat{\bm{m}}_{1}$ and 
$\hat{\bm{m}}_{2}$ are
two arbitrary unit vector along which magnetization is pointing, with the condition that 
$\hat{\bm{m}_{1}}$ $\neq$ $\hat{\bm{m}}_{2}$.  We show that these conductance variation as a function of absolute direction of magnetization is in phase with the polarization of outgoing current, therefore, it measures the non-equilibrium spin current. Moreover we show that to detect spin currents 
magnetization reversal is
not necessary, rather tilting the magnetization away from the original direction is
enough. Magnetization tilting can be achieved using external magnetic fields as is done
usually\cite{dieny}. The spin currents generated by SO interaction can also
tilt the magnetization because it will give rise to a torque on ferromagnetic detector. 
The torque arises because the spin current generated by SO interaction
in general will have components transverse to the magnetization direction of Ferromagnetic contact.
The transverse component of spin current is transferred to the FM as torque\cite{pareek75}.
If the torque is large enough it can tilt the magnetization.
For nm scale thick 
ferromagnetic film it is possible increase the torque by increasing the current density\cite{myers}. 
In case torque is not large enough to tilt
magnetization one can always use an external magnetic filed to rotate the magnetization.
Recently Adagideli, Bauer and Halprerin in Ref. \cite{bauer} have also discussed the possibility
of detecting torque related signal via small angle magnetization dynamics in three terminal systems. 
We stress that our proposal works in two terminal systems and relies on conductance measurements
which are easier as well have potential to be used in already developed semiconductor technology.

To study the above discussed effect quantitatively we make use of spin density matrix scattering theory
developed in Ref. \cite{pareek75} which
treats the non-magnetic and magnetic system 
at the same footing. An incident current in channel $|n,\alpha\rangle \equiv|n\rangle \otimes |\alpha\rangle$($n$ channel index due to transverse confinement along {\it y} direction and
$\alpha$ is spin index) leads to the following spin density matrix
for outgoing current\cite{pareek75},
\begin{equation}
\rho_{out}^{n\,\alpha}=\frac{1}{N}\sum_{m}\sum_{\beta\,\gamma}
T_{m\,n}^{\beta\,\alpha}T^{\dagger\,\alpha\,\gamma}_{n\,m}
|\beta\rangle\langle\gamma|.
\label{eq1}
\end{equation}
where N is normalization factor ensuring that $Tr(\rho)=1$ and $T_{m\,n}^{\beta\,\alpha}$ and 
$T^{\dagger\,\alpha\,\gamma}_{n\,m}$ are spin resolved transmission coefficient.
The incident current is unpolarized (completely mixed state) 
therefore, the
spin density matrix for the incident current 
is $\rho_{in}=n_{\alpha}|\alpha\rangle+n_{-\alpha}|-\alpha\rangle$ with 
$n_{\alpha}$=$n_{-\alpha}$.
Here $n_{\alpha}$ and $n_{-\alpha}$ is the number of up electrons and down electrons respectively.
The fact that incident current is unpolarized leads to the following
spin density matrix for outgoing current,
\begin{equation}
\rho_{f}=n_{\alpha}\sum_{n}\rho_{out}^{n\,\alpha}+n_{-\alpha}\sum_{n}\rho_{out}^{n\,-\alpha}.
\label{rhof}
\end{equation}

From eq.\ref{rhof} and eq.\ref{eq1} it is clear that spin
density matrix for outgoing current is completely determined if spin resolved transmission coefficients
are known. Since the polarization direction of outgoing current in general will
lie along any arbitrary direction, therefore, the transmission coefficient should be calculated in
eigen-basis of $\bm{\sigma} \cdot \hat{\bm{u}}$ where $\hat{\bm{u}}\equiv (\sin\theta \cos\phi,\sin\theta \sin\phi,\cos\theta)$ is the quantization axis. The polarization along the quantization axis $\hat{\bm u}$ is given by
the trace of matrix product of ${\bm \sigma} \cdot \hat{\bm u}$ and $\rho_{f}$ given in eq.\ref{rhof}.
However to determine polarization vector we need to calculate polarization components 
along two more direction which are perpendicular to $\hat{\bm u}$ and are linearly independent.
Therefore, 
for simplicity and clarity we choose two specific cases of quantization axis $\hat{\bm u}$, namely 
(a) We fix $\theta=\pi/2$
and vary $\phi$. In this case three linearly independent axis are $\hat{\bm{n}}_{1}=(\cos\phi,\sin\phi,0)$,
$\hat{\bm{n}}_{2}=(-\sin\phi,\cos\phi,0)$ and $\hat{\bm{n}}_{3}=(0,0,1)\equiv\hat{\bm{z}}$.
(b) We fix $\phi=0$ and vary $\theta$. In this case three linearly independent axis are $\hat{\bm{n}}_{1}=(\sin\theta,0,\cos\theta)$,
$\hat{\bm{n}}_{2}=(-\cos\theta,0,\sin\theta)$ and $\hat{\bm{n}}_{3}=(0,1,0)\equiv\hat{\bm{y}}$.
The polarization along a specific axis $\hat{\bm{n}}_{i}$ is calculated using, 
\begin{equation}
P_{i}=Tr[\rho_{f} (\bm{\sigma} \cdot \hat{\bm{n}}_{i}) ], \,\,\,\ i=1,2 ,3
\end{equation}
where $\rho_{f}$ is density matrix given by eq.\ref{rhof}. 

To obtain results we perform numerical simulation for a 
two dimensional electron gas (2DEG) using spin dependent recursive
Green-function technique. The details of this can be found in Ref. \cite{tribhu_pramana,tribhu}.
We model the conductor on a square tight binding lattice of size 50$\times$50
with lattice spacing {\it a} and hopping parameter {\it t}.
We fix the Fermi energy $E_{F}=1.0t$ and
take $t$=1 as the unit of energy. The tight binding dimensionless 
SO parameter is $\alpha=\lambda/(t\,a)$.
For disorder we take Anderson model,
with on-site energies distributed randomly between [-U/2, U/2],
where U is the width of distribution. 
The results presented are for
Rashba Spin-Orbit coupling \cite{rashba}. However the conclusion remain valid for other
type of SO couplings as well.

\begin{figure}
\includegraphics[width=\linewidth,height=1.9in,angle=0,keepaspectratio]{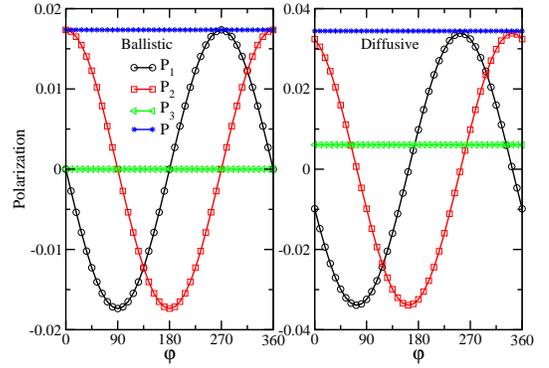}
\caption{\label{fig2}
(Color online) Spin polarization of outgoing current versus quantization axis
($\theta$=90 and $\phi$ is varied) for non-magnetic 
systems corresponding to fig.~1(a).
The dimensionless 
SO interaction 
strength $\alpha$=0.1 and $E_{F}$=1.0t and $k_{f}l=100a$($l$ is mean free path) for diffusive case. Straight line with filled circle
corresponds to the magnitude of polarization.}
\end{figure}

The three linearly independent polarization components of outgoing current
are shown in Fig.~2 for non-magnetic configuration shown in Fig.~1(a).
We have fixed $\theta=\pi/2$ and rotated the quantization axis in {\it xy} plane
which correspond to the case (a) discussed above.
For ballistic system (Fig.~2,left panel) outgoing currents 
is perfectly polarized along {\it y} axis (curve for $P_{2}$ has maximum value at
$\phi$=0,and $2\pi$ while $P_{1}$ has maximum and minimum values at $\phi=\pi/2$ and $3\pi/2$ respectively.) 
This is because current is flowing along positive {\it x} axis
the effective Rashba interaction becomes $ \lambda(k_{x}\sigma_{y})$ as 
mentioned in the
introduction.
However in the presence of disorder due to scattering the current direction in 2DEG changes randomly
therefore the Rashba coupling retains its two dimensional nature.
Hence all three components of polarization are non zero for diffusive system (Fig.~2, right panel).
Moreover the magnitude for each component as well the net polarization (shown by straight line with filled circles in Fig.~2)
is large in diffusive regime compared to the ballistic case.
This is because in presence of
disorder electron mobility decrease which in turn increase the effect of Rashba coupling.
Therefore moderate disorder will be helpful in generating spin current which is promising from the
experimental point since it will not require very clean or ballistic sample. 
\begin{figure}
\includegraphics[width=\linewidth,height=1.9in,angle=0,keepaspectratio]{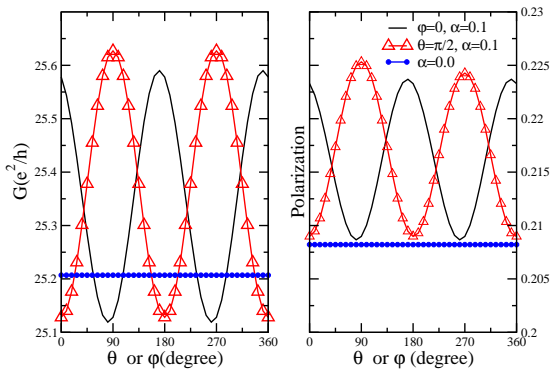}
\caption{\label{fig3}
(Color online) Conductance and Spin polarization of outgoing current as
a function quantization axis for ballistic system with ferromagnetic detector shown in Fig.~1(b).
Other parameters are same as in
fig. 2.}
\end{figure}

From Fig.~2 it is clear that the outgoing current is spin polarized though the
leads are non-magnetic (Fig.~1(a)).
Since the leads are non-magnetic
therefore the transmission coefficient will be independent of spin polarization
of outgoing current and as a consequence conductance will also be independent of quantization axis.
Hence with non-magnetic leads one cannot measure the spin current.
However, if the detector lead is magnetic (configuration shown in Fig.~1(b)) then transmission coefficient will become
spin dependent which will cause conductance to depend on quantization axis as discussed
in paragraph 3. This is confirmed in Fig.~3, where we have plotted the conductance(left panel) and 
polarization (right panel) for the configuration corresponding to Fig.~1(b).
Ferromagnet detector is modeled as exchange split with exchange splitting ($\Delta$)
given as $\Delta/E_{F}$=0.5 \cite{tribhu_pramana,tribhu}. 
The quantization axis
is given by the direction of magnetization. 
We have plotted results for two different quantization axes discussed in paragraph 4 above, namely,
(1) $\phi$ =0 and $\theta$ is varied, rotation in {\it xz} plane (2) $\theta$=90 and  $\phi$ have been varied, rotation {\it xy plane}.
We notice that conductance and polarization for zero SO coupling is independent of
quantization axis shown by straight line
with filled circles in Fig.~3. A finite polarization in absence of SO coupling is due to the presence of
ferromagnetic lead. In presence of SO interaction polarization increases compared to the polarization for zero SO coupling which means spin currents is being generated by SO coupling. 
Moreover polarization and conductance show in phase variation as a function of
$\theta$ or $\phi$. Therefore conductance measures the spin current as discussed before.
We notice that, the polarization curve in Fig.~3 for $\theta$=90 have different values for $\phi$=90 and $\phi$=270 which corresponds to {\it y} and {\it -y} axis respectively. 
This is because in ballistic system 
outgoing current is polarized along positive {\it y axis} as discussed 
earlier, therefore, the net polarization decreases (increases) when ferromagnet
points anti-parallel (parallel) to {\it y} axis. The asymmetry in polarization can be reduced
by reducing the magnetization of FM (decreasing $\Delta/E_{F}$) and by using a weakly coupled detector lead(non-invasive).
This would allow one to measure spin currents just due to SO coupling.
These finding are robust and survives in presence of disorder as shown in Fig.~4.
In presence of disorder 
due to scattering the outgoing current gets polarized along some arbitrary direction therefore the
asymmetry is visible throughout the $\theta$ or $\phi$ range. However the polarization variation still remain in phase with the conductance.
We stress that to measure spin current one does not need to reverse the magnetization because the conductance
changes as soon as the magnetization is tilted away from the original direction
as seen in Figs.~3 and 4.
\begin{figure}
\includegraphics[width=\linewidth,height=1.9in,angle=0,keepaspectratio]{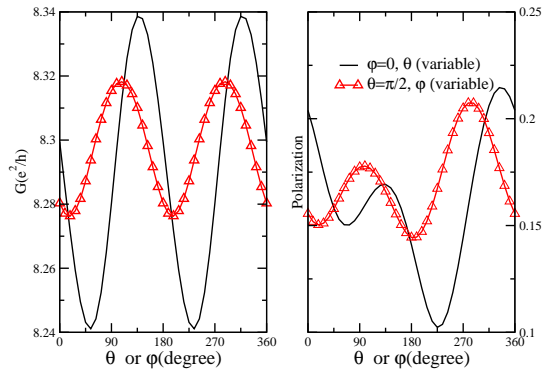}
\caption{\label{fig4}
(Color online) Conductance and Spin polarization of outgoing current
a function of quantization axis for diffusive system with ferromagnetic detector shown in Fig.~1(b). The results are for diffusive case with $k_{f}l=100a$($l$ is mean free path). Other parameters are same as in
fig.~3} 
\end{figure}

The magnetization can be tilted due to the
torque which arises because of spin currents as discussed in paragraph three. Dimensionless torque is defined as\cite{pareek75} ${\bm \tau}=\frac{{\bm M}\times {\bm P}}{{(\hbar\,I/\mid e \mid)}}$, where $\bm M$ is magnetization vector of
FM, $\bm P$ is
polarization vector of outgoing current and $I$ is current.
We have plotted the magnitude of torque in Fig.~5
for ballistic (left panel) and diffusive system(right panel) respectively. 
For ballistic system the torque goes to zero when magnetization is
either parallel or antiparallel to {\it y} axis (see Fig.~5 left panel, $\theta$=90 and $\phi$ variable)
because outgoing current is polarized along {\it y} axis as discussed above and
as a consequence torque vanishes.
For all other magnetization direction and for diffusive system (Fig.~5,  right panel), the torque is
always non zero. If the current density is large enough and Ferromagnetic film is of nm scale thick then,
magnetization can be tilted just by increasing current and as a consequence one would see a change in
conductance which measures the spin current as mentioned earlier.

\begin{figure}
\includegraphics[width=\linewidth,height=1.9in,angle=0,keepaspectratio]
{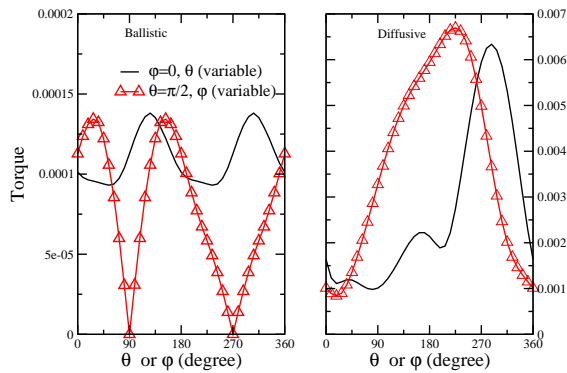}
\caption{\label{fig_cond} (Color online) Torque acting on detector ferromagnet as a function of quantization axis corresponding to configuration shown in Fig.~1(b).
}
\end{figure}

Finally let us discuss the possibility of observing the spin current related effects
presented above. From Figs.~3 and 4 it is seen that conductance changes by
2-3$\%$ when magnetization direction is rotated. Magneto-resistance experiments can 
easily detect changes of this order \cite{dieny}.
The 2-3$\%$ conductance change was obtained for
dimensionless Rashba SO parameter $\alpha$=0.1. This corresponds to the Rashba spin splitting
energy of the order of 10meV. The Rashba SO splitting of 30meV has been reported in Ref.\cite{molenkamp1}
in II-IV HgTe system. Therefore, the dependence of conductance on absolute
direction of magnetization can be easily detected which measures the spin currents.
In addition, tunability of Rashba coupling via external gate voltage\cite{molenkamp1} provides 
further control over
the spin current generation, torque and associated conductance change. 

In conclusion, we have presented a general method to generate and measure non-equilibrium spin currents
in two probe configuration via conductance. We have clarified that magnetization reversal is
not required to measure spin currents and shown that it is consistent with Onsager symmetry.
Further we have show that torque arising due to spin current can be used to tilt the
magnetization which will help in measuring spin currents.

\end{document}